
\input PHYZZX.TEX

\PHYSREV
\tolerance 2000
\nopubblock
\titlepage

\title{COUPLED BURGERS EQUATIONS - A MODEL OF POLYDISPERSIVE SEDIMENTATION}

{\author{Sergei E. Esipov}

\address{James Franck Institute and Department of Physics, University of
Chicago, 5640 South Ellis Avenue, Chicago, Illinois 60637, USA}
\vfil
\eject

\abstract{This paper compares theory and experiment for the kinetics of
time-dependent sedimentation. We discuss non-interacting
suspensions and colloids which may exhibit behavior similar to the
one-dimensional motion of compressible gas. The velocity of sedimentation
(or creaming) depends upon the volume fraction of the constituting particles
and leads to Burgers-like equations for concentration profiles. It is
shown that even the bi-dispersive system of two coupled Burgers equations
has rich dynamics. The study of polydispersive case reveals
a continuous ``renormalization'' of the polydispersity. We compare the Burgers
system evolution with the experimental results on mono- and polydispersive
sedimentation. The influence of thermal fluctuations is briefly discussed.}


\chapter{Introduction}

The study of the motion of particles in a fluid goes back to Einstein and
before that to Brown. Here we consider the effect of gravity upon the
particles. If they are heavier than the surrounding fluid the resulting motion
is called sedimentation; if lighter it is creaming. The conventional and
simplest problem involves the evolution of initially uniform
suspension or colloid.  At the bottom (of the test tube) there appears
the sediment, and at the top - water free of particles which is called
supernatant. Between these regions there is the original suspension itself.
Thus, two interfaces are created, and they spread and propagate
towards each other. The description of the bottom interface invokes high
volume ratios leading to ``traffic''-like problems with possible jams and
other instabilities whereby additional interfaces may emerge [1].
In this paper we address the evolution of the top
interface which we can study by thinking about the dilute limit, in which
the volume fraction of the particles is much less than 1.
We apply the continuity equation which
describes the conservation of species with concentration $c(x,t)$ and current
${\bf J}(x,t)$.
Such an equation is well-known in
the case of very small particles which experience Brownian motion. It can be
taken from Landau \& Lifshitz [2, \S 58,59]
$$\partial_t c + {\rm div} {\bf J} = 0\eqn\lanlif$$
where the particle flux ${\bf J}$ is produced by
external forces (such as gravity),
and also depends on gradients of concentration,
pressure, and temperature. In the experiments which we discuss below,
the influence of the (other than hydrostatic) pressure gradient and
temperature gradient is negligible, so that the particle flux can be written
as an expansion in the concentration gradient in the form
$${\bf J} = {\bf V}(c)\ c - D(c){\bf\nabla} c,\eqn\flux$$
where the sedimentation velocity ${\bf V}(c)$ is produced by gravity and the
hydrodynamic interactions, and the particle diffusivity $D(c)$ is produced both
by thermal (Brownian) motion and hydrodynamic motion.
Eqs\lanlif, \flux\ have {\it not} been applied
quantitatively to sedimenting systems. It is instructive to
discuss the possible reasons.

The kinetic coefficients
${\bf V}(c)$ and $D(c)$ are known for small enough particles when the Brownian
motion dominates. These mixtures are called colloids.
The velocity, ${\bf V}(c)$,
has been calculated in the profound paper by Batchelor
[3], see also [4].
Batchelor also found the dependence of Brownian diffusivity upon
concentration [5]. Batchelor's theory provides results for mono- and
polydispersive suspensions. Yet, the task of plugging these results into
Eqs\lanlif, \flux\ and comparing quantitatively
its evolution with real experiments
has not been performed. There have been
some qualitative results, though. It was recognized [6,7]
that Eqs\lanlif, \flux\
with $c$-dependent velocity resemble very closely the famous Burgers
equation of one-dimensional compressible flow [8,2]. This equation leads
naturally to shock waves in which the non-linear velocity dependence on
concentration is balanced
by diffusivity. We suggest here that the shock waves are the mathematical
realization of the observed interfaces. Some non-linear aspects of this
effect were studied long ago by Kynch [9], who discussed
the shock wave toppling in the absence of diffusivity,
and more recently by Barker and Grimson [6], and by van Saarloos and Huse [7].
Baker and Grimson pointed out the fact that the Burgers equation has
an analytical solution (if the concentration dependence of diffusivity is
neglected) and argued its qualitative applicability to a sedimentation
experiment [10].
No quantitative comparison was done.
The interesting
paper by van Saarloos and Huse [7] used Burgers equation to discuss the
sedimentation layers which are sometimes observed in the
course of sedimentation.
We shall discuss this phenomenon in a later paper [11].

Here we seek a comparison between
the time-dependent evolution predicted by Burgers equation and real
experimental data [12-14]. We shall also
extend the description of sedimentation
within the framework of Burgers equation to
suspensions with large non-Brownian particles. Some unresolved
theoretical issues arise in this area. For large enough particles
the Brownian diffusivity is surpassed by the so-called hydrodynamic
diffusivity. This effect has been studied by
different experimental groups [12,14-17] but has not been
fully understood theoretically, although a number of steps in this direction
has been made [18-20]. Interesting
numerical results have been reported by Ladd [21]
for model system consisting of 32 and 108 particles.
The value of the hydrodynamic diffusivity
depends upon unknown
density-density correlation function of a suspension. This correlation
function is not short-ranged as in the case of small enough particles
(colloids) - the property
used in the calculations of Batchelor [3,4] - and the dependence
$V(c)$ for suspensions is also unknown theoretically.
We avoid the unresolved theoretical issues by
making a simple order of magnitude estimate
of hydrodynamic diffusivity which is different from previous
estimates and consider the numerical
prefactor as a fitting parameter. We also ignore the tensor-like structure
of hydrodynamic diffusivity  which is hopefully insignificant
in the one-dimensional description offered by Eqs\lanlif, \flux\ for
concentation field depending solely upon vertical coordinate.
As for the velocity of hindered settling,
it will be desribed by an empirical
dependence.

We would like to go further in applying the Burgers equation.
This paper contains a straightforward generalization to the
systems of coupled Burgers equations which are capable of describing
realistic polydispersive suspensions. We find the steady-moving solutions
of coupled Burgers equations and reproduce the evolution of the suspension
interface discussed originally by Smith [23] and, in implicit form, by
Davis and Acrivos [24]. The coupled Burgers
equations predict an interesting phenomenon, which we named
{\it phase shifts}, it may be already observed
in a bidispersive system. Imagine that we study the evolution of intitially
homogenious bidispersive system. In a while there will be two steady-moving
interfaces, corresponding to the two given species. Their velocities
can be found by using Smith formulas [23]. What is the distance
between these interfaces? This distance cannot be found by multipying
the velocity difference by elapsed time. One has to account for the
stage of ``disentanglement'' of the interfaces interacting via the
non-linear coupled Burgers equations. Thus, the distance acquires the mentioned
phase shift which is of order of combined diffusive length calculated at the
disentanglement time, i.e. it is a diffusivity-dependent effect. This effect is
important for calculation of the width of interfaces of polydispersive
suspensions. On the other hand the quantitative definition of
phase shifts is not really straightforward: this phenomenon becomes important
when an experimentalist can spatially
resolve the interface profile, and possible definitions contain
a fair degree of uncertainlty. For example, one may expect a definition
of the distance between the interfaces as a measure between the points
where the concentrations of corresponding species diminish
in, say, $e$ times. However, in the
course of sedimentation in real polydispersive systems the segregation
of particles improves gradually with time, and it is unclear which
collection of species is to be traced with the above defintion. We then
regard the solution of the coupled Burgers equations as a necessary
step in describing the spreading of the interface.
In the case of real polydispersive suspensions
the distribution function of particle sizes $a$,
$c(a,x,t)$ replaces the concentration. This distribution function
evolves in a complex way near the interface. Step by step the system eliminates
particles with different $a$. We call this evolution
{\it renormalization} since it is produced by applying many times a given rule
of eliminating the fastest species
(infinitely many times in the continuum limit).

We introduce the Burgers equation in Section 2 and discuss the
relevant kinetic coefficients.
To show how this equation works, the
experimental data of Ref.13 for a colloid
are compared with the analytical solution. In Section 3 we used coupled
Burgers equations to describe polydispersive systems. We
first address the unique steady-state
motion it terms of simple analytical formulas borrowed from physics of
one-dimensional compressible gas [2] and then simulate
the bi-dispersive case numerically to investigate the evolution
of the initial condition problem.
Description of the continuous polydispersive case
is presented in Section 4. In the polydispersive case sedimentation
leads to the continuous renormalization of the particle size distribution
function as one scans through the top interface. The late-time solution is
discussed together with the effects of diffusivity. Comparison with
experiments [12,14] is given.
Section 5 contains
brief discussion of influence of intrinsic
thermal fluctuations on the sedimenting system and indicates the connection
with dynamical critical phenomena.

\chapter{Monodispersive sedimentation}

Depending on the particle size one distinguishes suspensions and colloids
for a given fluid. The boundary is defined somewhat arbitrarily through
a P\'eclet number,
$$Pe = {2aV_0\over{D_0}} = {8\pi{g}\Delta\rho{a}^4\over{3k_bT}},\eqn\pe$$
which describes competition between action of (say) gravity and thermal
fluctuations on a spherical particle of radius $a$. Gravity results in
particle downward motion with the Stokes velocity
$V_0$,
$$ V_0 = {(4/3)\pi a^3 g
\Delta\rho\over{6\pi\eta{a}}} = {2 a^2 g
\Delta\rho\over{9\eta}},\eqn\velstokes$$
due to the density difference $\Delta\rho$ in a fluid with
with viscosity $\eta$ in the presence of gravity acceleration $g$.
Thermal fluctuations
are the source of the Brownian diffusivity,
$$\quad D_0 = {k_B T\over{6\pi\eta{a}}},\eqn\Brown$$
where $k_BT$ is the temperature in energy units.
Systems with $Pe<1$ are conventionally
called colloids. Thermal fluctuations are strong for colloids and
lead to a uniform density-density
correlation function which is short-ranged. In the opposite case
the structure and kinetic properties
are exclusively determined by the interparticle hydrodynamic interactions.
For water at
room temperature
and particles less that $1\mu{m}$ Brownian
effects usually dominate, while larger particles form suspensions.

Evolution of the interface which separates the region of the initial
distribution and supernatant in colloids
can be described by the concentration profile $c(x,t)$ which
obeys a Burgers-like equation (see above)
$$\partial_t c = \partial_x [V(c)c] + \partial_x [D(c)\partial_x c],\eqn\diff$$
where $V(c), D(c)$ represent the Stokes velocity and gradient diffusivity
modified by the presence of other
particles. The axis $x$ is directed upward, opposite to the direction of
gravity.
If $c(x,t)$ is normalized to be the volume fraction ($c={4\over3}
\pi{a}^3n$ with $n$ being the number density), then
$$V(c) = V_0 f_v(c),\quad ,\eqn\vel$$
where $f_v(c)$ is the hindering effect and
$$f_v(c) = 1 - k c + O(c^2),\qquad c\ll1.\eqn\hind$$
According to calculations by Batchelor [3],
$k\approx 6.55$ in the dilute limit, $c\ll 1$.
Larger values of the concentration are sometimes approximated, for example, by
the Richardson-Zaki empirical formula [22], $f_v(c)=(1-c/c_0)^{kc_0}$.
(There exist other suggested formulas [13].)
Here $c_0 < 1$ is the volume fraction at dense packing.

The diffusivity $D$ entering \diff\ can be written as
$$D_b(c) = D_0 f_b(c),\eqn\ditil$$
with (again according to Batchelor [5], see also [25])
$$f_b(c) = 1 -(k-8)c + ...\approx 1 + 1.45 c, \quad c\ll1. \eqn\hintil$$
The nonlinear terms in the expansion \hintil\
(studied experimentally [25]) lead to a maximum of $f_b$
at $c\sim 0.15$, so that the entire concentration dependence of the diffusivity
is within 10\% of its bare values up to the volume fractions $c\sim 0.3$.

Addressing now suspensions,
we recall that the right-hand side of Eq\diff\ represents a
double expansion of the particle current in terms of small concentration
value and small spatial gradients of concentration.
In the case of large P\'eclet numbers,
we deal with strongly interacting
systems, and the applicability of Eq\diff\ is unclear. To be more specific
we do not doubt that the term ${\bf V}(c)c$ must enter the particle flux
${\bf J}$ in the homogenious suspension,
see Eq\flux, however the existence of a local (and/or fickian)
expression for the inhomogeneity-induced part of flux has not been established.
In fact, the numerical simulations by Ladd [21] indicate that motion
of a marked particle is somewhat non-gaussian.
Nevetheless, in this paper we use Eqs\flux\ and \diff\ to
model the evolution of suspensions. Then, the function $D(c)$ is
conventionally called hydrodynamic diffusivity [19].
This concept is an attempt to get the main effect of multiparticle
forces without a very detailed calculation and can only be justified
{\it a posteriori}, as it is sometimes the case with the Fokker-Planck
method.

Thus, the hydrodynamic diffusivity of suspensions,
which will be denoted as $D_h(c)$, reflects the net effect of the
velocity fluctuations due to interparticle hydrodynamic interactions.
This problem is reminiscent of those which arise in the description of
dense plasma [20]. However, hydrodynamic forces
decay with particle-particle separation like $1/r$ and provide
even stronger interaction as compared to $1/r^2$ Coulomb forces.
For Coulomb forces it is known (\S 27,41 in Ref.26) that plasma experiences
collective
and chaotic motions, the latter due to interparticle collisions.
Hydrodynamic collective modes are dissipative, so the interesting
question is the account for the collisions. Collisions in 3D plasma that
are of most importance correspond to the small angle scattering events
(\S41, Ref.26, see also [27]).
In the course of these events particles
approach each other at a distance of order of the Debye screening
length although the interparticle separation is much smaller than that.
Mean ``free'' path is inversely proportional to the
volume fraction $c$, $l\sim 1/na^2 \sim a/c$. Diffusivity is
the product of the mean velocity fluctuations, $v$ and $l$, $D(c)\sim av/c$.
This is the $c$-dependence which is valid for usual gases (\S 7,8, Ref.26),
interacting plasma (\S 43, Ref.26), and
it was claimed by Koch and Shaqfeh [20] for hydrodynamic diffusion in the
dilute limit.

Unfortunately, the pair collision approximation, which was developed by
Landau for plasma (\S 41, Ref.26) does not work here:
for the particles with equal densities and of identical size there
is a well-known reversibility and symmetry (see, for example, the
concise discussion by Hinch [19]).
Namely, the solution for
viscous flow surrounding two descending spheres provides {\it no}
hydrodynamical repulsion and/or attraction, and the inter-particle
separation remains constant in time.
Thus, at least triple collisions are needed, and a question immediately arises
whether it is legitimate to use triple collisions and disregard high
order ones? Analogous situations do arise in plasma [28].

Koch and Shaqfeh argued that a particle distribution
without correlations is a solution of the Liouville continuity equations
descibing all particles in a point-particle approximation. They continued
that finite particle sizes are important to destroy this (say, initial) uniform
distribution. Then Koch and Shaqfeh concluded that the most important triple
collisions are those when two out of three particles are within $O(a)$
distance while the third particle is far away. We do not see any reason
for such a conclusion. This conclusion also implies that there exists a way
to resolve the question of higher order collisions without identifying
a small parameter different from $c \ll 1$. In plasma physics this additional
parameter is the condition of high temperature,
$T \gg e^2/\bar{r}$ (\S 27, Ref.26).
Namely, the plasma temperature must greatly
exceed the Coulomb interaction at the interparticle separation $\bar{r}$.
This condition ensures that the Debye radius greatly exceeds $\bar{r}$,
justifies the Landau collision integral and so on. At present we do not think
it is possible to extend these ideas to the physics of sedimentation.
Interestingly, the
numerical results by Ladd [21] do indicate some screening, although it is
quite different from what is presented in [20].

Hinch [19] pointed an argument by Caflish that uniform distribution
with the finite particle sizes leads to convective flows due to density
fluctuations. Hinch speculated that the convection will cease at the
level of mixing which is comparable to the
interparticle separation, $ac^{-1/3}$. The velocity fluctuations
of the ``remaining'' convection are then of the order of the main term
of the interparticle force, $V_0c^{1/3}$. We may continue this speculation by
multiplying these two values,
$$D_h(c) = k'aV_0, \qquad c\ll1.\eqn\hyddif$$
where $k'$ is the pure number.
One may expect that in the zero concentration limit the effect of
hydrodynamic diffusivity vanishes. This does not necessarily imply that
$D_h(c)$ must go to zero, since the required system size
and time needed to establish the steady-state regime both
diverge as the concentration goes to zero.
Using \hyddif\ one finds that the ratio $D_h/D_b$
is of the order of the P\'eclet number. If the dependence on
concentration in Eq\hyddif\ is incorrect, there would exist an additional
class of mixtures which cannot be ascribed to either suspensions or
colloids.

Experimentalists have been analyzing hydrodynamic diffusion by two
different techniques. One approach [12,14] is to investigate the
gradient diffusion. The other is to trace a marked particle and
evaluate the self-diffusivity [15-17]. In a strongly interacting system
the two approaches may, in principle, produce different values.
Ham and Homsy [15] found that for the case of self-diffusion the factor
$k'$ depends upon concentration for small
$c < 0.01$, and saturates at about $k'=5$ for larger values of $c$.
Experiments by Nicolai {\it et al} discriminate between vertical and
horizontal self-diffusivities and indicate $k'=6-10$ for the former.
There is a possibility that the whole tensor may be inferred from
experimental data by  Nicolai {\it et al}.  As for the gradient
diffusion analysis it will be discussed in detail below.
We shall estimate $k'=10$ from analyzing
gradient diffusion of
the experiments [12,14] (see Section 4). This estimate is within the
error bars of the
experimental estimates of self-diffusivity. The $c$-dependence of $D_h$ was
not really tested in this paper since both the degree of polydispersity and
uncertainty of the size distribution function used in [12,14]
are found to be
too large for this purpose.

We have introduced the Burgers equation and corresponding kinetic
coefficients and, before we begin to use it, let us recall two
approximations which are usually made when deriving equations \diff-\hintil.

1. The particles are supposed not to interact in any fashion
apart for the hydrodynamic interaction and excluded volume effect, i.e.
these particles are hard spheres. Such an approximation
rules out many interesting systems exhibiting spinodal decomposition
and phase interface formation [29].

2. Thermal fluctuations are neglected in Eq\diff\ except for the mean-field
contribution to diffusivity $D_b$. These are ``three-dimensional''
fluctuations which have to be averaged over the lateral plane for each
height $x$ of the sedimenting system. Alternatively, one can consider
the 3D-version of Eq\diff\ with anisotropic drift in the vertical direction.
As discussed by van Saarloss and Huse [7], the conservative noise leads only
to small corrections (see also Section 5).

Now we attempt a detailed comparison with the experiemnt. It is well-known [8]
that for any smooth initial distribution $c(x,t)$ with the boundary conditions
$c(-\infty,t)=c_0$, $c(\infty,t)=0$ the late-time asymptotic is
a single Burgers shock moving with a constant speed and width. Indeed,
substituting $c(x,t) = c(x+vt)$ into \diff\ and integrating from $x$ to
$\infty$
one obtains
$$D(c){dc\over{dx}} = [v-V(c)]c,\eqn\once$$
which for large negative $x$ gives $v=V(c_0)$. The shape of the interface
can be found by another integration of Eq\once. This shape depends
upon P\'eclet number.
Using \vel-\hintil\ we find an implicit equation for the
concentration dependence on $x$
$$ \Big({8\over{k}} - 1\Big)\log(c_0-c)
-{1\over{kc_0}}\log\Big({c_0-c\over{c}}\Big) = {x Pe\over{2a}}.
\eqn\twice$$

At intermediate times the experimentally resolved interface profile is
time-dependent. We found relevant measurements in a paper by
Al-Naafa and Selim [13] who studied
the evolution of the
optical density of a monodispersive colloid.
To make the comparison, we solve Eq\diff\ with initial condition
$c(x,0)=c_0\theta(x)$ to get the evolution of concentration
$$c(x,t) = {c_0f_1(x,t)\over{f_2(x,t)+f_1(x,t)}},$$
$$f_1(x,t) = \exp\Big\{-{V_0kc_0\over{D_0}}[x+V_0t(1-kc_0)]\Big\}\Big\{
1 - {\rm erf}\Big[{x+V_0t(1-2kc_0)\over{2\sqrt{D_0t}}}\Big]\Big\},$$
$$f_2(x,t) = 1 + {\rm erf}\Big[{x+V_0t\over{2\sqrt{D_0t}}}\Big].\eqn\sol$$
We then try to fit the
data in Fig.5, paper [13] using this formula. The result is shown in Fig.1.
The experimental parameters are given in the figure caption.
Note that there are {\it no} adjustable parameters used.
Initially the shape of the optical density profile is more narrow than the
prediction of formula \sol. After a transient we get the optical shape
which does fit.

We do not have any explanation of the transient.
Our attempt to use more complex fit
(see Section 4 for the description of polydispersive fit)
by taking into account the small polydispersity of the suspension ($\pm 4.5$
nm)
did not result in significant changes. Thus the experiment both supports and
denies the validity of Burgers equation for sedimentation. We need
more experimental data to help resolve this issue.

\chapter{Bidispersive sedimentation}

We then consider the
case of bimodal distribution of particle sizes. Let $c_1$ and $c_2$ be the
concentrations of these particles, both obeying continuity equations
\lanlif. The corresponding fluxes $J_1, J_2$ are
$$J_i = V_i(c_1,c_2) - D_i(c_1,c_2){\partial
c_i\over{\partial x}}.\eqn\cfluxes$$
Using expansion at small concetrations we arrive to the two coupled Burgers
equations
$$\partial_t c_1 = V_1\partial_x[(1-k_{11}c_1-k_{12}c_2)c_1] +
D_1\partial_{x}^2c_1$$
$$\partial_t c_2 = V_2\partial_x[(1-k_{21}c_1-k_{22}c_2)c_2] +
D_2\partial_{x}^2c_2.\eqn\couple$$
This system is a straightforward generalization of the monodispersive
Burgers equation \diff. The hindering velocity of each of the particle
species depends linearly on both volume fractions. This dependence
has been studied by Batchelor and Wen, see Ref.4 where one can find
method of computing the
constants $k_{11}, k_{12}, k_{21}, k_{22}$.
The dependence of diffusivities $D_1, D_2$ upon particle concentrations
will be ignored in this Section together with the cross-terms when
a gradient of, say, $c_1$ influences a flux of $c_2$ particles. These
refinements (all vanishing with $c$)
can wait until the validity of Burgers equation is understood.

First, we are looking for steady-moving solutions $c_{1,2}(x-v_{1,2}t)$.
Substituting these into equations \couple\ and integrating once we have
$$D_1 {dc_1\over{dx}} = (v_1-V_1)c_1 +
V_1k_{11}c_1^2 - V_1 k_{12}c_1c_2\eqn\interm$$
and an analogous second equation. Two shocks may be observed
for the relevant boundary conditions $c_{1,2}(-\infty)=c_{01,02},
c_{1,2}(\infty)=0$.
We then have to decide which of the waves is faster. Without loss of
generality let us assume that $v_1 > v_2$ where
$$v_1 = V_1(1-k_{11}c_{01}+k_{12}c_{02}),\eqn\first$$
$$v_2 = V_2(1-k_{21}c_{01}+k_{22}c_{02}),\eqn\secondfirst$$
This leads to the first shock velocity to be $v_1$.
The substance 2 changes its concentration when the first shock passes through.
Let us denote this changed (we shall also call it renormalized)
concentration as $c_2^*$. At the top of substance 1 we essentially have the
problem of monodispersive sedimentation considered in previous Section.
Let the hindered velocity of settling be $v_2^*$, as we have seen above,
it is also the velocity of the second shock.
Concentration $c_2^*$ obeys the quadratic continuity equation
in the reference frame moving with the first shock velocity $v_1$
(see Ref.2, Chapter IX, and Ref.23)
$$c_2(v_2-v_1) = c_2^*(v_2^*-v_1),\eqn\square$$
here $v_2^*$ is
$$v_2^* = V_2(1-k_{22}c_2^*).\eqn\second$$
Fig.2 illustrates the steady-state geometry of the system.
It can be easily shown that if $v_1 > v_2$ then
$v_1 > v_2^*$, and $v_1\rightarrow v_2^*$ when $v_1\rightarrow v_2$.
Consequently, no other steady-moving solutions exist for this system.
Thus, the steady-moving solution of coupled Burgers system are characterized by
the well-known formulas for the interface (shock) velocities [23,25,30].
It is nice to get to know that these
formulas also describe the solutions of coupled Burgers equations.
Experimental confirmation of shock velocities predicted by Eqs\square,
\second\ for bidispersive (and tridispersive)
systems can be found in Refs.25,30.

The coupled Burgers equations also allow to study the transient
regimes prior to the establishing of the steady-moving shocks.
Transient regimes were not reported experimentally, most probably
due to the absence of relevant theoretical method of
analysis, and we have no available experimental data to fit the
concentration profiles. We hope to encourage
studies of time-dependent phenomena by addressing physics of coupled Burgers
equations. To give an example of transient behavior we solved the system
of two coupled Burgers equations numerically.

The two shocks when formed separate
linearly in time and their asymptotical
shapes may be found by numerical integration of
Eq\interm\ with real concentration dependences of the diffusivities, see
[5].
Both shocks have finite widths (analytical formulas may be obtained in
a number of limits) and quickly cease to interact, see Fig.2. It must be
emphasized that no matter how simple the steady-moving solution of the
system \couple\ may seem, the transition to this solution in time may be
rich, see Fig.3 where a change in coupling constant $k_{21}$ from 0.35 to 0.4
resulted in system
inability to reach the dynamical steady-state within the integration time.
Figs 2 and 3 show examples of the time dependent
solutions  of Eqs\couple\ which develop transient raises of concentration
and even additional transient shocks (Fig.3). The final profiles
in Fig.2 display the two regions discused above: first shock at $x\approx30$,
accompanying with renormalization of $c_2$ from 1 up to 1.86; and the
second shock at $x\approx45$ at the end of the integration time.
Measuring rate of change of the separation between adjacent profiles
one can verify that the shock velocity is nicely predicted by Eqs\first,
\second. However, the separation between the shocks accounts
for the initial condition and interaction at the
stage of shock ``disentanglement'' as well as for the widths of shocks.

It is also noteworthy that the described behavior is just an example
of dynamics of a compressible one-dimensional gases.

\chapter {Polydispersive sedimentation}

It is not hard to apply the ideas of previous Section
to the real polydispersive case. We shall first discuss the situation
where finite number $N$ of particle sizes is involved and particle
groups are numbered, $1 \leq i \leq N$. As before
one should calculate the hindered velocities of sedimentation of all
particle groups which are given by
$$v_i = V_i\Big[1 - \sum_{j=1}^{j=N} k_{ij}c_j\Big],\eqn\many$$
and represent a generalization of expressions like \first. Again
we refer to Ref.14 for the constants $k_{ij}$. The maximal velocity
${\max}_i(v_i)$
is to be found and the corresponding substance (say, $i_0$)
will form the lowest (fastest) shock.
Generally speaking, the motion of the edge of the size distribution
is not necessarily the fastest, one may rather say that {\it some}
substance $i_0$ will be move down
first, depending upon the matrix $k_{i,j}$ and
size (and density) distribution function. Other particle concentrations
at the top of the leader will be
renormalized, according to the equations
$$v_i^* = V_i\Big[1 - \sum_{j=1}^{j=N} k_{ij}c_j^*\Big],
\eqn\manyr$$
$$c_i(v_i-v_{i_0}) = c_i^*(v_i^*-v_{i_0}),\eqn\contin$$
which form a system of quadratic equations. Note, that the system can be
formally extended to $i,j=i_0$ given that $c_{i_0}^* = 0$.

Now the entire procedure is to be repeated $N$ times. At each level of
renormalization one additional substance is eliminated and others get
renormalized. An example can be seen in Fig.4 for the
evolution of initially Gaussian-like size distribution function with $N=26$
different particle sizes, and real parameters taken from paper [12].
In this particular case there is no inversion of
ordering, i.e. larger sizes are eliminated first. The result of the
calculation is a sequence of successively eliminated particle species
and corresponding velocities. The interface shape becomes wider in time,
although not as wide as one would obtain without renormalization.
The concentration profile in the limit of vanishing diffusivity
can be found by adding regions with renormalized concentrations
at separations prescribed by successive velocity differences.
Renormalization leads to
thinner interfaces, this is the so-called
phenomenon of {\it self-sharpening} of the interface [12,14]
when smaller particles at the top of larger particles move faster then
they would do if mixed with larger particles. When diffusivity is finite - a
contribution
to the interface width comes from the phase shifts.

Before moving on to the continuous case we discuss the influence of
diffusivity from yet a different angle.
For a single Burgers shock connecting concentration
change from $c_{-\infty}$ to $c_{+\infty}$ the width is given by
$$W = {2D\over{V_0k(c_{-\infty}-c_{+\infty})}}.\eqn\wid$$
where for a moment we assume the simplest possible situation:
constant diffusivity and monodispersive case.
The smaller is the concentration difference the weaker is the non-linear
effect competing with constant diffusivity [8],
and the shock width becomes proportionally larger. Similar phenomenon
exists in the polydispersive case, the larger is the number of different
particle sizes which are taken into consideration - the smaller
are the concentration changes, the smaller are the velocity differences.
In addition, the widths of the successive shocks overlap.
Thus, diffusivity provides
a time-dependent cut-off for the achieved resolution in particle size step.

This physics is reflected in the renormalization of size
distribution function. When particles of a given size move down
and are not present at some height, the distribution function of this size
at that height becomes zero
(rigorously speaking it is, of course,
of order of exponentially small, $exp(-Vx/D)$, diffusive corrections).
In the
limit of large $N$ this leads to a discontinuity
which moves monotonously or jumps depending
on absence or presence of ordering inversion discussed above. Discontinuous
renormalization of the size distribution
function implies that diffusivity is to be taken into account, and
the existence of the diffusive time-dependent cut-off regularizes the problem.
At any point along the
interface new structure appears as time goes on. The minimal resolved size
step diminishes with time, and the
emerging of the fine structure of the
distribution function (or segregation) continues. Note that certain
integrated properties such as the profile of the optical
density may be still perfectly smooth under these circumstances, it is the
size distribution function which changes most drastically within the
suspension interface.

In the continuous case the coupled Burgers equations form an
integro-differential equation for the evolution of the size distribution
function $c(a,x,t)$
$${\partial{c(a,x,t)}\over{\partial{t}}} = V_0(a){\partial\over{\partial{x}}}
\Big[c(a,x,t) - \int_0^{\infty}da' k(a,a')c(a,x,t)c(a',x,t)\Big] +$$
$$
{\partial\over{\partial{x}}}\int_0^{\infty}da'
D(a,a'){\partial{c}(a',x,t)\over{\partial{x}}}
\Big),
\eqn\inb$$
where kernel $k(a,a')$ is the continuous generalization of matrix $k_{ij}$
and diffusivity is a functional of $c$ with diagonal terms of zero and
first order in $c$ and off-diagonal terms of first order in $c$.
The solution of Eq\inb\ is well-defined due to the presence of $D$.
It is noteworthy that for the kernels $k(a,a'), D(a,a')$ which dependence
on $a$ and $a'$ consists of factorizable terms it may be more convenient to
work with the Mellin
transform of Eq\inb\ performed in variable $a$.

Generalization of Eqs\manyr, \contin\ to the continuous case is
straightforward. We consider the simplest case of no ordering inversion.
Then at any point $x$, on the interface (at late times) there exists
maximal particle size $b(x)$ which is ``eliminated'' at this point.
It is convenient to
parametrize the spatial dependence by $b(x)$. Then concentration becomes
a function of $a$ and $b$, i.e. $c(a,b)$ (and certainly an implicit function of
$x$ through $b(x)$).
Particle hindered velocities are given by
$$v(a,b) = V_0(a)\gamma(a,b),\quad \gamma(a,b) = 1 - \int_0^b da' k(a,a')c(a')
,\eqn\infr$$
and the continuity equation reads
$$c(a,b)[v(a,b)-v(b,b)] = c(a,b-db)[v(a,b-db)-v(b,b)],\eqn\infc$$
where
$$v(a,b-db) = V_0(a)\Big\{1 -
\int_0^{b-db}da' k(a,a')c(a,b-db)\Big\}.\eqn\rv$$
Expansion of  \infr\ - \rv\ to the first order to get an equation for $\partial
{c}(a,b)/\partial{b}$ should be done carefully
since $c(a,b)$ contains a singularity
at $a=b$
$$c(a,b) = {C(b)\over{(b-a)^{\mu(b)}}} + c^R(a,b),\eqn\extr$$
where $\mu$ is positive, such that integral \infr\ converges, and $c^R(a,b)$
stands for the regular part at $a=b$. The exponent $\mu(b)$ can be
related to other functions by using Eq\infc, \rv.
A study not presented here shows that
$$\mu(b) = {{\partial\ln\gamma(a,b)\over{\partial{b}}}\over{{2\over{b}}+{
\partial\ln\gamma(a,b)\over{\partial{a}}}}}_{| a=b},\eqn\gam$$
i.e. concentration $c(a,b)$ becomes infinite at the point
$a=b$. Given that the singularity is identified, an explicit
differential equations of renormalization
can be written for $dC/db$ and $\partial c^R(a,b)/\partial{b}$.

Numerical integration of Eq\inb\ enables to study time-dependent evolution of
polydispersive suspensions. For this purpose we performed simulations
on a Sun Sparc 2 workstation to fit the experimental data by Davis and
Hassen [12], and Lee {\it et al} [14].
The simplest possible explicit difference scheme (Euler scheme)
already works nicely for Eq\inb, provided that spatial and temporal
steps obey the conditions $\Delta{x}\ll D/V_0$,
$\Delta{t}\ll\min(\Delta{x}/V_0,\Delta{x}^2/D)$. The results
cease to depend on numerical resolution for $N>30$ particle
species (to model the size distribution function) and more than
1000 spatial points
for each of the species given that the characteristic value of the
hydrodynamic diffusivity exceeds $10^{-4} cm^2/sec$.
Diffusivity of particles of size $a$
was selected to be
$$D_h = k'a_0V_0(a_0),\eqn\polydiff$$
where $a_0$ is the average size.
The results are presented in Figs. 4, 5 and 6. Fig.4 shows evolution
of the size distribution function $c$ and resembles closely the
numerical results by Davis and Hassen (see their Fig.2) if the latter are
resolved in space. One can clearly see the appearance of singularities of
the size distribution function. Their amplitude is restricted by
diffusive resolution as we discussed above. Fig.5
compares simulation results with the
experimental data for transmitted light intensity, which is given by [12]
$$\log[I(x,t)] \propto - \int {da\over{a}}c(a,x,t),\eqn\lighti$$
scaled to the experimental amplitude range $I(x,0)/I(x,\infty)=0.12$.
The factor $k'$ was the adjustable parameter. It is interesting
to note that we didn't find any
significant dependence upon the constant $k'$ for the values $k'$ up to
about 10. Also important is the observation that the interface width
obtained at smaller $k'$ (we tried $1 \leq k'\leq 10$)
is nicely comparable with the experimentally
measured one. Smaller values of $k'$ require higher resolution or
more advanced numerical schemes and were not attempted.

Our conclusion is different from the conclusion made by Davis and Hassen
who used Eqs\manyr, \contin\ and discovered that in addition to the
polydispersive width there is a diffusive-like contribution which they
identified with the effect of gradient
hydrodynamic diffusivity. We agree with this
statement but point out that ``bare'' hydrodynamic diffusivity that enters
Eqs\diff, \inb\  doest not contribute to the interface width
directly - it is achieved through the evolution of the equation.
For instance, the distribution of the phase shifts influences the width.
Thus, we think that the simulation of the coupled Burgers equation \inb\
is useful to decribe the results of the
polydispersive sedimentation experiment.
It should also be noted that the tails of the
interface profile is sensitive to the real
particle size distribution (which is not necessarily gaussian as assumed in
Fig.5 and in Ref.12).

An analogous fit was performed for the data in Fig.2 of Ref.14. The fit
is shown in Fig.6.
Here the experimental standard deviation of particle size was only
2/3 of that of Ref.[12], and polydispersity alone can not account for
the observed width. With the help of hydrodynamic diffusivity
(using $k'=10$) we can get curves which are rather similar to experimental
ones.
The exception is the last curve, which position is not described by the
motion with a constant velocity. We are thankful to Soonchil Lee who
explained that the shape and position of this curve is affected by the
presence of the lower interface between the suspension and sediment, and
must be ignored in our analysis [31].

\chapter{Thermal fluctuations}

Now we discuss briefly the influence of intrinsic thermal noise on
sedimentation which
corresponds to Brownian motion of sedimenting particles and is only
relevant at small P\'eclet numbers. One can further discriminate
between two situations. In macroscopic fluid mechanics equilibrium thermal
fluctuations have zero correlation length. Correspondingly, different particles
experience different Brownian drag with amplitude reflecting their sizes.
One may consider adding a delta-correlated (in size, space and time)
gaussian noise $\xi(a,{\bf x},t)$ to the 3D generalization of
Eq\inb (c.f. [7]).
On the other side sufficiently small colloidal particles may be within
the correlation of some thermal (especially Van-der Waals)
fluctuations, then there is a correlated
Brownian drag on particles of different sizes, and additive noise
is no longer delta-correlated in particle size. This leads to complex
behavior: for each noise-induced inhomogeneity a segregation of particles
starts within initially uniform system. In the same manner as usual
Burgers turbulence is an interplay between non-linearity and diffusion,
we have here in addition the effect of size-coupling.

Even for the monodispersive system in thermal equilibrium with average
concentration $c_0$ at temperature $T$ there are fluctuations, $c_1=c-c_0$,
which are described by Burgers equation,
$$\partial_t c_1 = V_0\partial_x [(1-2kc_0-kc_1)c_1] +
D\Delta{c_1} + \xi.\eqn\diffluc$$
Here we again neglected the $c_1$-dependence of diffusivity, $\xi$ is the
thermal noise with the correlator proportional to temperature.
If one averages Eq\diffluc\ over the lateral cross-section of the
sedimentation column, $\langle...\rangle_{\perp}$,
the resulting equation is no longer a closed Burgers
equation but their mean-field approximations coincide. The value of
$h(x) = \int^x \langle{c_1(x,t)}\rangle_{\perp}$
is then analogous to the so-called interface height
which was extensively studied in the framework of
Burgers turbulence [32] and represents an example of dynamical critical
phenomena. Nonequilibrium dynamics can
be induced by changing the ambient temperature (liquid helium?)
or switching off the interparticle interaction.

Another question is the influence of external thermal fluctuations when
experiment is performed under incomplete thermostatic conditions.
This issue was addressed experimentally [33] and discussed
theoretically [7]. The detailed study of this effect which
results are different from Ref.7 will be
published elsewhere [11].

\chapter{Conclusion}

Application of (coupled) Burgers equation to describe
sedimentation of (polydispersive) suspension and colloids enables
one to reproduce experimental data and provides a basis for
analyzing time-dependent sedimentation.
We hope that this procedure will become a routine lab software and will
help developing an industrial analysis for sedimentation processes.

\ack {I am grateful to David Grier, Daniel Mueth and John Crocker for
attracting my attention to sedimentation dynamics,
and to Leo Kadanoff for his constant interest, discussions and great
help with the manuscript. I also benefited from a conversation with Elisabeth
Guazzelli who also provided me with Refs. 16-21.
Numerical simulations of the two
coupled Burgers equations were made by Michael Brenner, his
participation is gratefully acknowledged. The work was
supported in part by the MRSEC Program of the National Science Foundation
under Award Number DMR-9400379
and in part by NSF Grant NSF-DMR-90-15791.}
\chapter{References}
\item{1.} Y. Zimmels, Powder Technology, {\bf 70}, 109, (1992)
\item{2.} L.D. Landau and E.M. Lifshitz, {\sl Fluid Mechanics}, Pergamon
Press, NY, 1975
\item{3.} G.K. Batchelor, J. Fluid Mech., {\bf 52}, 245, (1972)
\item{4.} G.K. Batchelor, J. Fluid Mech., {\bf 119}, 379, (1982);
G.K. Batchelor, C.-S. Wen, J. Fluid Mech., {\bf 124}, 495, (1982);
J.M. Revay and J.J.L. Higdon, J. Fluid Mech., {\bf 243}, 15, (1992)
\item{5.} G.K. Batchelor, J. Fluid Mech., {\bf 74}, 1, (1976)
\item{6.} G.O. Barker and M.J. Grimson, J. Phys. A: Math. Gen., {\bf 20}, 305,
(1987)
\item{7.} W. van Saarloos and D. Huse, Europhys. Lett., {\bf 11}, 107, (1990)
\item{8.} The basic references on Burgers equations are:
J. M. Burgers, {\em The Non-linear Diffusion Equation }, Reidel,
Dordrecht, (1974); G.B. Witham, {\sl Linear and Nonlinear Waves}, New-York:
Wiley, (1974)
\item{9.} G.J. Kynch, Trans. Faraday Soc., {\bf 48}, 166, (1952)
\item{10.} M.-C. Anselmet, R. Anthore, X. Auvray, C. Petipas and R. Blanc, C.R.
Acad. Sc. Paris, {\bf 300}, S\'erie II, 69, (1985)
\item{11.} T. Dupont, S.Esipov and P. Hosoi, unpublished.
\item{12.} R.H. Davis and M.A. Hassen, J. Fluid Mech., {\bf 196}, 107, (1988)
\item{13.} M.A. Al-Naafa and M.S. Selim, AIChE Journal, {\bf 38}, 1618, (1992)
\item{14.} S. Lee, Y. Jang, C. Choi, and T. Lee, Phys. Fluids A, {\bf 4}, 2601,
(1992)
\item{15.} J.M. Ham and G.M. Homsy, Int. J. Multiphase Flow {\bf 14}, 533,
(1988)
\item{16.} H. Nicolai, B. Herzhaft, L. Oger, E. Guazzelli, and E.J. Hinch,
Phys. Fluids {\bf 7}, ???, (1995)
\item{17.}  H. Nicolai and E. Guazzelli, Phys. Fluids, {\bf 6}, ???, (1995)
\item{18.} R.E. Caflisch and J.H.C. Luke, Phys. Fluids, {\bf 28}, 759, (1985)
\item{19.} E.J. Hinch, ``Sedimentation of small particles'', in {\it Disorder
and Mixing}, edited by E. Guyon, J.-P. Nadal and Y. Pomeau, Kluver
Academic Publishers, Nato ASI Series E: Applied Sciences, vol.152, Dordrecht,
(1988)
\item{20.} D.L. Koch and E.S.G. Shaqfeh, J. Fluid Mech., {\bf 224}, 275, (1991)
\item{21.} A.J.C. Ladd, Phys. Fluids A, {\bf 5}, 299, (1993)
\item{22.} J.F. Richardson and W.N. Zaki,
Trans. Inst. Chem. Engrs., {\bf 32}, 35, (1954)
\item{23.} T.N. Smith, Trans. Inst. Chem. Engrs., {\bf 44}, 153, (1966)
\item{24.} R.H. Davis and A. Acrivos, Ann. Rev. Fluid Mech., {\bf 17}, 91,
(1985)
\item{25.} M.A. Al-Naafa and M.S. Selim, Fluid Phase Equilibria, {\bf 88}, 227,
(1993)
\item{26.} E.M. Lifshitz and L.P. Pitaevskii, {\sl Physical Kinetics}, Pergamon
Press, New York, (1981)
\item{27.} S.E.Esipov, I.B.Levinson, Adv. in Phys. {\bf 36}, 331, (1987)
\item{28.} S.E.Esipov, I.A.Larkin, Sov. Phys. JETP {\bf 62}(6), 1200,
(1985)
\item{29.} See, for example:
T. Biben, J.-P. Hansen , and J.-L. Barrat, J. Chem. Phys., {\bf 98},
7330, (1993)
\item{30.} R.H. Davis and K.H. Birdsell, AIChE Journal, {\bf 34}, 123, (1988)
\item{31.} Soonchil Lee, private communication.
\item{32.} J. Krug and H. Spohn, in {\it
Solids far from Equilibrium: Growth, Morphology and Defects}, edited by
C. Godreche (Cambridge University Press, Cambridge) 1990.
\item{33.} See D.B. Siano,
Journal of Colloid and Surface Science, {\bf 68}, 111,
(1979) and references therein. Analogous experiments are under way in the
David Grier's group at the University of Chicago.

\chapter{Figure Captions}

\item{Fig.1} Fit of light intensity data versus tube height
for a monodispersive suspension [13].
Curve 1 was taken after 195.2 hours after the
beginning of experiment, curve 2  - after 374.7 hours. Experimental
parameters used for the fit: $a=65 nm$,
$c_0=0.009$, $\Delta\rho=1.005\ g.cm^3$,
$V_0=0.12\times10^{-5} cm/sec$, $T=25^o C$, so that
$\eta=0.0077\ g/cm sec$, $D_0=4.4\times10^{-8}
cm^2/sec$. Both curves are scaled in vertical
direction to meet experimental range of Fig.5 in Ref.13.
This is the experiment at small P\'eclet numbers, so that
hydrodynamic diffusivity is negligible with respect to the Brownian part.
\item{Fig.2} Numerical integration of two coupled Burgers equations \couple.
Parameters: $V_1=0.1$, $V_2=0.05$, $k_{11}=k_{12}=k_{22}=0.4$, $k_{21}=0.35$,
$D_1=D_2=0.01$.
Initial conditions for $c_1(x)$ and $c_2(x)$ were selected to be their
individual Burgers shocks in the absence of coupling with small
shift with respect to each other, $c_{-\infty}=1$, $c_{+\infty}=0$ for
both substances. Integration time $t=2000$, $\Delta{t}=40$
between profiles. $c_1$ is shown with lines, $c_2$ with broken lines.
System size $x=100$. The sign of velocities was selected such that the
shock motion occurs from right to left. Initial and final
profiles are shown with thick lines. Eqs\couple\ do not impose any
contraints on amplitudes of $c_{1,2}$. We can
formally use the non-physical range $c_{1,2} > 1.$
\item{Fig.3} An example of complex temporal behavior. Parameters are the same
as in Fig.2 except for $k_{21}=0.4$.
\item{Fig.4} Snapshot of the solution $c(a,x,t)$ at time $t=36\ min$ for the
experiment [12]. Parameters: $a=61 (\pm 6)\ \mu{m}$, $\Delta\rho=1.384\
g/cm^3$, $c_0=0.05$, $\eta=0.0085
g/cm sec$. With this parameters one obtains: $V_0(a)=0.0141\ cm/sec$,
$D_h(a)=k'\times 8.5\  10^{-5}
cm^2/sec$. The value $k'=1$ is used (see Fig.5).
System length is 40 cm, initially gaussian size distribution with
deviation $6 \mu{m}$ is used. Number of species $N=26$,
1600 spatial points for each particle species.
\item{Fig.5} Fit of raw transmitted light intensity versus time for a
polydispersive suspension (Fig.11 of Ref.12). The intensity is measured
at heights $x=1.5, 6.0, 24, 32\ cm$. Note that we use
$k=2.5$; this is the experimental value (!) which can be
easily extracted from Fig.11 in contradiction with Table 1
(both from Ref.12). The other possible
explanation is that Fig.11 was, in fact, taken for $c_0=0.02$, then
$k\sim 5$ in agreement with Fig.12 of Ref.12 and Table 1. (In both
cases the conclusion made in the text remains valid).
The Richardson-Zaki
dependence $(1-c)^k$ is employed instead of $1-kc$, here $c$ is the total
volume fraction. The value $k'=1$ is used to show that the hydrodynamic
diffusivity is not required to account for the width of curves.
At the top of theoretical
curves one can notice small oscillations caused by insufficiently large
number of particle species $N=26$ when they get separated.
\item{Fig.6} Fit of volume fraction $\int da\ c(a,x,t)$ versus height $x$ for
four different times: 332, 443, 567, and 1288 $sec$ from left to right.
Last experimental curve at 2611 $sec$
is not fitted since its average displacement
is not described by a motion with constant velocity (see text).
Experimental curves are from Fig.2 of Ref.3. Parameters: $a=67.9 (\pm 4.0)\
\mu{m}$, $\Delta\rho=1.57\ g/cm^3$, $\eta = 0.0164\ g/cm sec$, $c_0=0.1$.
With these parameters:
$V_0=0.0096\ cm/sec$, $D_h=k'\times 6.5 10^{-5} cm^2/sec$.
The value $k`=10$ is used. The Richardson-Zaki
dependence $(1-c)^k$ is employed with $c$ being the total volume fraction
here.

\end